\documentclass[preprint,5p]{elsarticle}
\journal{Expert Systems with Applications}

\usepackage{graphicx}
\usepackage{multirow,array,tabularx}
\graphicspath{{./figures/}}
\DeclareGraphicsExtensions{.eps,.jpg}

\usepackage{amsmath}
\usepackage[caption=false,font=footnotesize]{subfig}
\usepackage{amssymb}
\usepackage{url}

\begin{document}
	
\begin{frontmatter}

\title{Resolving Congestions in the Air Traffic Management Domain via Multiagent Reinforcement Learning Methods}
\author[1]{Theocharis Kravaris} 
\author[2]{Christos Spatharis} 
\author[1]{Alevizos Bastas}
\author[1]{George A. Vouros} 
\author[2]{Konstantinos Blekas}
\author[3]{Gennady Andrienko} 
\author[3]{Natalia Andrienko}
\author[4]{and Jose Manuel Cordero Garcia}
\address[1]{University of Piraeus, Piraeus, Greece e-mail: georgev@unipi.gr}
\address[2]{Dept. of Computer Science \& Engineering, University of Ioannina, Ioannina, Greece e-mail: kblekas@cs.uoi.gr}
\address[3]{Frauhofer Institute IAIS, Sankt Augustin, Germany e-mail: \{gennady.andrienko, natalia.andrienko\}@iais.frauhofer.de }
\address[4]{CRIDA, Spain e-mail: jmcordero@crida.es}

\begin{abstract}
In this article, we report on the efficiency and effectiveness of multiagent reinforcement learning methods (MARL) for the computation of flight delays to resolve congestion problems in the Air Traffic Management (ATM) domain. Specifically, we aim to resolve cases where demand of airspace use exceeds capacity (demand-capacity problems), via imposing ground delays to flights at the pre-tactical stage of operations (i.e. few days to few hours before operation). Casting this into the multiagent domain, agents, representing flights, need to decide on own delays w.r.t. own preferences, having no information about others' payoffs, preferences and constraints, while they plan to execute their trajectories jointly with others, adhering to operational constraints. Specifically, we formalize the problem as a multiagent Markov Decision Process (MA-MDP) and we show that it can be considered as a Markov game in which interacting agents need to reach an equilibrium: What makes the problem more interesting is the dynamic setting in which agents operate, which is also due to the unforeseen, emergent effects of their decisions in the whole system.  We propose collaborative multiagent reinforcement learning methods to resolve demand-capacity imbalances: Extensive experimental study on real-world cases, shows the potential of the proposed approaches in resolving problems, while advanced visualizations provide detailed views towards understanding the quality of solutions provided.
\end{abstract}


\end{frontmatter}

\section{Introduction}
Congestion problems, modelling situations where multiple agents demand to use resources of a specific capacity simultaneously, exceeding resources’ capacity, are ever present in the modern world. Most notably, congestion problems are typically very complex and large-scale and appear regularly in various real-life domains (urban traffic congestion, air traffic management and network routing). Consequently, it is of no surprise that they have drawn much attention in the AI and autonomous agents research (e.g. \cite{Agogino2012},\cite{Bazzan2009},\cite{Kuyer2008},\cite{Tumer2007},\cite{Walraven2016}) for at least two decades \cite{Dresner2004} and have been the focus of game theoretic models for much longer \cite{Rosenthal1973}\cite{Milchtaich2004}.

In the air-traffic management (ATM) domain, congestion problems arise naturally whenever demand of airspace use exceeds capacity, resulting to {\em “hotspots”}. This is known as the {\em Demand - Capacity Balance (DCB) problem}. Hotspots are resolved via airspace management or flow management solutions, including regulations that generate delays and unforeseen effects for the entire system, increasing the factors of uncertainty regarding the scheduling of operations. For instance, they cause the introduction/increase of time buffers in operations’ schedules and may accumulate demand for resources within specific periods per day. These effects present further multiple negative effects to ATM stakeholders and are also translated into costs and loss of reliability, including customers’ satisfaction and environmental effects.  Today, delays are imposed to flights without considering the propagated effects to the entire ATM system (e.g. to other flights and airspaces), which is inherently highly complex and dynamic. While delays may be due to several reasons, the high share is allocated to the increased demand for airspace use (over 90\% in some airspaces) \cite{EurocontrolPR2017}. It got significantly worse in 2018 \cite{EurocontrolPR2018} when delays across Europe more than doubled, due to the increase in traffic among other factors. In general, all performance analysis and studies lead to the idea that the ATM system is very close to, or already at, a saturation level. These issues, in conjunction to the forecasted increase in air traffic (e.g.  Eurocontrol as Network Manager forecasts increases in traffic of +50\% in 2035 compared to 2017, meaning 16 million flights across Europe \cite{Eurocontrol2018}\cite{Eurocontrol2018v2}) impose the need for the assessment and minimization of delays at the {\em “pre-tactical”} phase of operations (i.e. from several days to few hours before operations), also considering the effects of delays to the overall ATM system and the highly dynamic environment in which airspace users operate.

Indeed, resolving hotspots at the {\em pre-tactical} phase of operations and assessing delays early enough, support increasing the predictability of the overall system, alleviating many of the negative effects. One of the major objectives here is to minimize {\em ground delays} while ensuring efficient utilisation of airspace and fair distribution of delays among flights. The reduction of the average ground delay per flight for 1 minute, means vast cost savings for all ATM stakeholders. Therefore, in this paper we consider only {\em ground delays} and subsequently we succinctly call these {\em “delays”}.

To resolve the DCB problem, this article formalises the problem as a multiagent Markov Decision Process (MA-MDP), where agents, representing flights, aim to decide on own {\em ground delays}, jointly with others, with respect to own preferences and operational constraints on the use of airspace, while possessing no information about the preferences and payoffs of others: As said, the goal is to detect and resolve {\em all} hotspots at the pre-tactical phase of operations, considering the {\em joint} and propagated effects of agents’ ground delays to the evolution of airspace demand, so as to minimise delay costs, while ensuring efficient utilisation of airspace and fair distribution of  ground delays among flights, i.e. without penalizing a small number of them.

We show that this problem can be considered as a Markov game, in which interacting agents need to reach an equilibrium to conflicting delay preferences, while resolving hotspots in which they participate. As part of the formulation, we devise a reward function that considers agents' contribution to hotspots and implied cost when agents deviate from their schedule. To solve the problem, we propose multiagent Reinforcement Learning (MARL) methods, whose efficiency and efficacy is evaluated in real-world DCB problem cases, each one comprising flight plans for a specific day (i.e. 24 hours) above Spain. The data sources include real-world operational data regarding flight plans submitted just before take-off per day of operation, data regarding changing sector configurations per day of operation, and reference values for the cost of strategic delay to European airlines, currently used by SESAR 2020 Industrial Research \cite{Cook2015}. Details are provided in Section 2.

The agent-based paradigm introduced in this paper is in contrast to regulating flights in a first-come-first-regulated basis - as it is the case today in ATM: Regulations are imposed to airspaces, resulting to delays for flights entering that airspace using a first-come-first-delayed rule, without considering the implications of these delays to other flights operating in different airspaces and/or time periods.

A major conclusion of this article is that collaborative MARL methods reduce the average delay per flight quite effectively, managing to provide solutions to DCB problems, thus, imposing delays that result to zero hotspots. Indeed, results are quite significant, since in most of the cases the average delay per flight (i.e. the ratio of summing all delays to the total number of flights) is reduced considerably compared to the solutions provided by the Network Management organization (NM), while a small percentage of flights have been imposed delay more than half an hour, and only a small percentage of flights get delay.

We envisage the work laid out in this paper to be seen as a first step towards devising multiagent methods for deciding on delay policies for correlated aircraft trajectories, answering the call of ATM domain for a transition to a Trajectory Based Operations (TBO) paradigm (SESAR\footnote{\url{https://ec.europa.eu/transport/modes/air/sesar_en}} in Europe  and Next Gen in the US\footnote{\url{https://www.faa.gov/nextgen/}}). 

The contributions made in this paper are as follows:
\begin{itemize}
\item The Demand-Capacity Balance problem is formulated as a multiagent Markov Decision Process (MA-MDP). We also show that equivalently, it can be considered as a Markov game in which interacting agents need to reach an equilibrium. The problem formulation takes into account the dynamics of the real-world setting.
\item All methods are evaluated in real-world cases comprising large number of flights in busy days above Spain.
\item We show the effectiveness of multiagent reinforcement learning (MARL) methods to provide solutions to the DCB problem.
\item We show how visualizations of solutions allow to understand the merits and limitations of methods, proving a first level of explaining solutions and the rationale behind these. 
\end{itemize} 

Our previous works \cite{Kravaris2017, Spatharis2018a,Spatharis2018b}, reported on the potential of alternative multi-agent reinforcement learning methods, providing initial results. The differences between this article and our previous publications are as follows:
\begin{itemize}
\item In this article we provide a problem formulation as a Markov game,
\item We focus on the efficacy of a collaborative multi-agent reinforcement learning method, in comparison to independent reinforcement learning agents, showing also the tolerance of the method to incorporate preferences of agents, without reducing the quality of solutions,
\item The properties of the reward function proposed are discussed in comparison to those of other rewards used in resolving congestion problems,
\item An extensive literature review is provided, showing the exact contributions of our work,
\item Results from extensive number of experiments are reported in a more systematic way, providing results' statistical analysis,
\item We show how visualisations can advise on the quality of solutions, delving into the details of methods' strengths and limitations. 
\end{itemize}

The structure of this paper is as follows: Section 2 provides a specification for the DCB problem and introduces terminology from the ATM domain. In section 3 we give the related work on reinforcement learning techniques in resolving congestion problems. Section 4 presents the problem formulation within an MA-MDP framework and shows that the problem can be considered as a Markov Game. Section 5 then presents the reinforcement learning methods proposed for solving the problem. Section 6 presents the formulation of the reward function and discusses desired properties from that function. Section 7 presents evaluation cases and results. Finally, section 8 concludes the article outlining future research directions. The article is complemented by two Appendices: The first presents the methodology for exploiting data sources towards the construction of evaluation cases, and the second explores the capacity of the multiagent reinforcement learning methods to incorporate into their solutions airspace users’ constraints on delays.

\section{Problem Specification}
The current Air Traffic Management (ATM) system leads to congestion problems, casted as demand-capacity balance (DCB) problems, i.e. cases where imbalances regarding the demand of airspace use and the provided airspace capacity do occur.  With the aim of overcoming ATM system drawbacks, different initiatives, notably SESAR in Europe and Next Gen in the US have promoted the transformation of the current ATM paradigm towards a new, {\em trajectory-based} operations (TBO) one: The trajectory becomes the cornerstone upon which all the ATM capabilities will rely on. 

Flight trajectories cannot be considered in isolation from the overall ATM system: Intertwined operational aspects and factors of uncertainty introduced, lead to inefficiencies to trajectory planning and huge inaccuracies to assessing trajectory execution. Accounting for network effects and their implications on the joint execution of individual flights, requires considering interactions among trajectories, in conjunction to considering dynamic operational conditions that influence any flight. 

Being able to devise methods that capture aspects of that complexity and take the relevant information into account, would greatly improve planning and decision-making abilities in the ATM domain. 

Towards this goal, our specific aim is to assess ground delays that need to be imposed to planned flight trajectories before the actual operations, towards resolving DCB problems well-in-advance, considering ATM system dynamics and network effects due to interactions among trajectories.

\begin{figure*}
\centering
\includegraphics[width=0.4\linewidth]{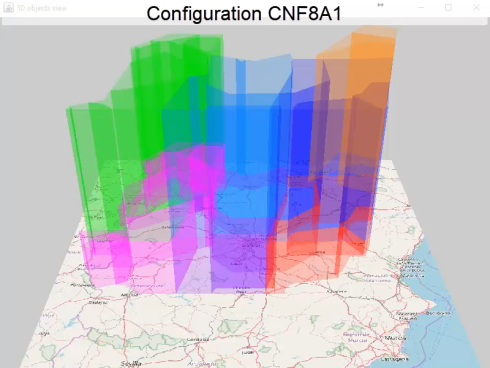} 
\includegraphics[width=0.4\linewidth]{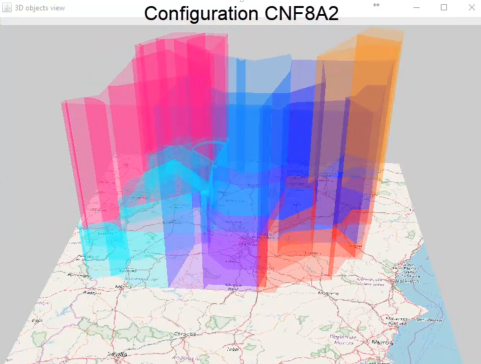}
\caption{Two of the configurations of sectors in the Spanish airspace. Colours are for distinguishing between sectors. Illustrations have been created using the V-Analytics platform \cite{Andrienko2013}.}
\label{fig:configurations}
\end{figure*}

More specifically, the DCB problem (or DCB process) considers two important types of objects in the ATM system: {\em aircraft trajectories} and {\em airspace sectors}. Sectors are air volumes segregating the airspace, each defined as a group of airblocks. Airblocks are specified by a geometry (the perimeter of their projection on earth) and their lowest and highest altitudes. Airspace sectorization may be done in different ways, depending on sectors’ configuration determining the active (open) sectors, as Fig.\ref{fig:configurations} shows. Only one sector configuration can be active at a time. Airspace sectorization changes frequently during the day, given different operational conditions and needs. This happens transparently for flights.

The {\em capacity} of sectors is of utmost importance: this quantity determines the maximum number of flights flying within a sector during any time period of specific duration (typically, in 60' periods).

The {\em demand} for each sector is the quantity that specifies the number of flights that co-occur during a time period within a sector. The duration of any such period is equal to the duration of the period used for defining capacity. 

There are different types of measures to monitor the demand evolution, with the most common ones being the {\em Hourly Entry Count} and the {\em Occupancy Count}. In this work we consider {\em Hourly Entry Count}, as this is the one used by the Network Manager (NM) at the pre-tactical phase. 

The {\em Hourly Entry Count (HEC)} for a given sector is defined as the number of flights entering the sector during a time period, referred to as an {\em Entry Counting Period} (or simply, {\em counting period}). HEC is defined to give a “picture” of the entry traffic, taken at every time “step” value along a period of fixed duration: The step value defines the time difference between two consecutive counting periods. For example, for a 20 minutes step value and a 60 minutes duration value, entry counts correspond to pictures taken every 20 minutes, over a total duration of 60 minutes.

{\em Aircraft trajectories} are series of spatio-temporal points of the generic form ($long_i$, $lat_i$, $alt_i$, $t_i$), denoting the longitude, latitude and altitude, respectively, of the aircraft at a specific time point $t_i$. Casting them into a DCB resolution setting, trajectories may be seen as time series of events specifying the entry and exit 3D points, and the entry and exit times for the sectors crossed, or the time that the flight will fly over specific sectors. Thus, given that each trajectory is a sequence of timed positions in airspace, this sequence can be exploited to compute the series of sectors that each flight crosses, together with the entry and exit time for each of these sectors.

Specifically, let us consider a finite set of discrete air sectors \textbf{R}=$\{R_1, R_2, ...\}$ segregating the airspace. As already pointed out, sectors are related to operational constraints associated to their capacity, whose violation results to demand-capacity imbalances: These are cases where $D_{R,p} > C_R,$ where {\em p} is a counting period of pre-defined duration {\em d}, $D_{R,p}$ is the demand for sector {\em R} during counting period {\em p}, and $C_R$ is the capacity of the sector for any period of duration {\em d} (equal to the counting period duration), in which the sector is {\em “open”}. 

Thus, a trajectory {\em T} in \textbf{T} is a time series of elements of the form:
\begin{equation}
T=\{ (R_1, entry_1, exit_1) .... (R_m, entry_m, exit_m) \},
\end{equation}
where $R_l, l=1 ,..., m$, is a sector in the airspace and {\em entry/exit} are time points of entering/exiting that sector.
This information per trajectory suffices to measure the demand for each of the sectors $R \in$ \textbf{R} in the airspace, in any counting period {\em p}. 
Specifically, the demand in sector R in period {\em p} is $D_{R,p}=|\mathbf{T}_{R,p}|$, i.e. the number of trajectories in $\mathbf{T}_{R,p}$, where:
\begin{center}
$
\mathbf{T}_{R,p}=\{T \in \mathbf{T} | T=(…,(R, entry_t , exit_t),…), $ 
{\em and the temporal interval }  
$[entry_t,exit_t]$  {\em overlaps with period $p$} $\}$.
\end{center}
Trajectories requiring the use of a sector {\em R} at the same period {\em p} causing a congestion, i.e. trajectories in $\mathbf{T}_{R,p}$ s.t. $D_{R,p} > C_R,$ are defined to be interacting trajectories for {\em p} and {\em R}.

We assume a trajectory-based operations environment, with an enhanced accuracy of pre-tactical flight information – provided by airlines’ flight plans. Pre-tactical flow management is applied prior to the day of operations and consists of planning and coordination activities. This operational environment is close to the one existing today, but it requires airlines to specify their flight plans during the pre-tactical phase, allowing the detection of hotspots based on planned trajectories and airspace operational constraints. While today the resolution of hotspots is done either by the Network Manager at the pre-tactical phase, or by the Air Traffic Controller at the tactical phase of operations, we aim towards their resolution at the {\em pre-tactical} phase. Resolving DCB problems at the pre-tactical phase of operations implies an iterative, collaborative process among stakeholders, during which flight plans submitted may change even just before take-off: The methods presented here pave the way to such a collaboration, through automation, but this is not within the scope of this work.

In addition, our efforts complement research on trajectory planning and accurate prediction of trajectories, as well as efforts towards enhanced information sharing abilities between stakeholders. Results from these orthogonal to our domain areas would further enhance the DCB process at the pre-tactical phase of operations and are out of the scope of our current work. 

In this operational context we consider an agent $A_i$ to be the aircraft performing a planned flight trajectory, in a specific date and time. Thus, we consider that agents and trajectories coincide, and we may interchangeably speak of agents $A_i$, trajectories $T_i$, flights, or agents $A_i$ executing trajectories $T_i$. Agents, as it will be specified, have own interests and preferences, and take autonomous decisions on resolving hotspots: It must be noted that agents do not have communication and monitoring restrictions, given that hotspots are resolved at the pre-tactical phase, rather than during operation.

To resolve hotspots, agents have several degrees of freedom: They may either change their trajectory to cross sectors other than the congested ones, or change the schedule of crossing sectors in terms of changing the entry and exit time for each of the crossed sectors. In this paper we consider only changing the schedule of crossing sectors by imposing {\em ground delays}: i.e., shifting the whole trajectory by a specific amount of time.

Now, the problem is about agents to decide on their delays so as to execute their trajectories jointly, in an efficient and safe way, w.r.t. sectors’ capacities. Specifically, in the DCB problem the goal is to:

\begin{itemize}
\item Resolve all demand-capacity imbalances, providing a solution with zero hotspots, in conjunction to 
\item minimizing the average delay per flight (ratio of total delay to the number of flights); so as to
\item distribute delays among flights without penalising a small number of them, and
\item utilise efficiently the airspace so as to have an even distribution of demand to sectors in all counting periods within a total period of trajectories’ execution \textbf{H}.
\end{itemize}

To resolve a hotspot occurring in period {\em p} and sector {\em R}, a subset of interacting trajectories in $\mathbf{T}_{R,p}$ must be delayed. It must be noted that agents have conflicting preferences towards resolving hotspots, since they prefer to impose the smallest delay possible (preferably none) to their own trajectory (i.e. at least one of them should have a greater delay than others, else the hotspot occurs later in time), or they may have different requirements on the maximum delay to be imposed to their flights. In any case they do need to execute their planned trajectories safely and efficiently. 

Clearly, imposing delays to trajectories may cause the emergence of hotspots to another time period for the same and/or other sectors crossed. This is due to the fact that the sets of interacting trajectories in different periods and sectors may change. This can be done in many different ways when imposing delays to flights, resulting to a dynamic setting for any of the agents. Thus, the sets of interacting trajectories do change unpredictably for the individual agents, according to agents' decisions and according to the changes in sectors’ configurations. 

Agents executing interacting trajectories and contributing to hotspots are considered to be “interdependent”, given that the decision of one of them directly affects the others. These dependencies provide a way to take advantage of the spatial and temporal sparsity of the problem: For instance, a flight crossing the northwest part of Spain in the morning, will never interact in any {\em direct} manner with a flight crossing the southeast part of the Iberian Peninsula at any time, or with an evening flight that crosses the northwest part of Spain. However, a flight may {\em indirectly} affect any other flight, due to ATM “network effects”. In addition, as mentioned above, dependencies between trajectories must be dynamically updated when delays are imposed to flights, given that trajectories that did not interact prior to any delay may result to be interacting when delays are imposed, and vice-versa. 

The dynamic {\em society} of agents $(\mathbf{A, E^t})$ is modelled as a dynamic {\em coordination graph} \cite{Guestrin2002} with one vertex per agent $A_i$ in \textbf{A} and any edge $(A_i, A_j)$ in $\mathbf{E^t}$ connecting agents with interacting trajectories in \textbf{T}, at time {\em t}. The set of edges are dynamically updated when the set of interacting pairs of trajectories changes.

$\mathbf{N^t}(A_i)$ denotes the {\em neighbourhood} of agent $A_i$ in the society, i.e. the set of agents interacting with agent $A_i$ at time instant {\em t} and in sector {\em R}, including also itself.

The {\em ground delay} options available in the inventory of any agent $A_i$ for contributing to the resolution of hotspots may differ between agents: These, for agent $A_i$ are in $\mathbf{D_i}=\{0,1,2,..., MaxDelay_i\}$. We consider that these options may be ordered by the preference of agent $A_i$ to any of them, according to the function $p_i: \mathbf{D_i}\rightarrow \Re$. We do not assume that agents in $\mathbf{A}-\{A_i\}$ have any information about $p_i$. This represents the situation where airlines set own options and preferences for delays, even in different own flights, depending on operational circumstances, goals and constraints. We expect that the order of preferences should be decreasing from 0 to $MaxDelay_i$, although, with a different pace/degree for different agents.

\subsection{Problem statement (Multiagent DCB problem resolution}

Considering any pair of interacting agents $A_i$ and $A_j$ in the society $(\mathbf{A, E^t})$, with $A_j$ in $\mathbf{N^t}(A_i)-\{A_i\}$, they must select among the sets of available options $\mathbf{D_i}$ and $\mathbf{D_j}$ respectively, so as to increase their expected payoff w.r.t. their preferences $p_i$ and $p_j$ (thus, minimizing flights’ delays): A solution consists of assignment of delays to flights, such that all imbalances are resolved, resulting to zero hotspots.

This problem specification emphasises on the following problem aspects: 
\begin{itemize}
\item Agents (i.e. individual flights) need to coordinate their strategies (i.e. chosen options to delays) to execute their trajectories jointly with others, considering traffic and network effects, w.r.t. their preferences and operational constraints;
\item Agents need to jointly explore and discover how different combinations of delays affect the joint performance of their trajectories, given that the way different trajectories do interact is not known beforehand: This is true, given that agents do not know in advance (a) the interacting trajectories that emerge due to own decisions and decisions of others, (b) the emergence of sectors’ open configurations, and (c) they do not know whether trajectories crossing sectors in new/emerging airspace configurations result to new hotspots; 
\item Agents' preferences and constraints on the options available may vary depending on the trajectory performed, and are kept private;
\item There are multiple and interdependent hotspots that occur in the total period \textbf{H} and agents have to resolve them jointly;
\item The setting is highly dynamic given that the agents’ society, the occurring hotspots and the sector configurations change unpredictably for individual agents.
\end{itemize}

\section{Related Work}
In this section we consider prior work related to (a) the resolution of the DCB problem, (b) the use of reinforcement learning techniques in resolving congestion problems and computing equilibria in coordination games, and (c) reward functions and their properties.

A comprehensive review of mathematical modelling and various formulations of demand-capacity imbalance problem is presented in \cite{Agustin2010}. This work reviews methods addressing congestions due to excess of the airport arrival and departure capacities, or of the airspace sector capacity. While most of early work refers to the simplest models, which do not consider airspace sectors, a category of methods addressing the Air Traffic Flow Management Problem attempts to solve real situations, also considering the airspace sector capacity. Additionally, while ground and en-route delays are important measures studied towards resolving congestions, methods addressing the Air Traffic Flow Management Rerouting Problem consider also the case where the flights can be diverted to alternative routes. As the authors point out, the problem becomes more realistic when changes in capacity are considered, which has led to incorporating stochastic methodologies for possible unforeseen changes. These methods focus mostly on the tactical phase of operations, rather on the pre-tactical.

More recent work has shown the importance and potential of multiagent reinforcement learning methods to address congestion problems in Air Traffic Management at the tactical level \cite{Agogino2012}\cite{Agogino2005}\cite{Crespo2012}\cite{Cruciol2013}\cite{Tumer2007}. This provides a shift from the current ATM paradigm, which rely on a centralized, hierarchical process, where decisions are based on flow projections ranging from one to six hours, resulting to slow reactions to developing conditions, potentially causing minor local delays to cascade into large regional congestion.  

The potential of reinforcement learning methods (either centralized or multiagent methods) to congestion problems, other than those in the aviation domain (e.g. to urban traffic) has received much attention in the recent years, with the most challenging issue being the coordination among agents, so as the solutions to increase agents’ individual payoff, in conjunction to increasing the whole system utility. Towards this target there are several proposals, among which the use of coordination graphs \cite{Kuyer2008}, where agents coordinate their actions only with those whose tasks somehow interact. 

The use of coordination graphs, where agents connected in pairs have to decide on joint policies, connects the computation of joint policies to computing equilibria in Markov games between interacting agents. Towards this goal, studies (e.g.\cite{Mukherjee2008}\cite{Sugawara2014}\cite{Yu2013}), have shown that Q-learners are competent to learners using for instance WoLF \cite{Bowling2002}, Fictitious Play \cite{Fudenberg1998}, Highest Cumulative Reward -based \cite{Shoham1997} models. Based on these conclusions, going beyond the state of the art and providing evidence on the potential of collaborative reinforcement learning methods to compute social conventions in complex settings, the work in \cite{Vouros2017} proposes social Q-learning methods, according to which agents interact with all of their acquaintances, considering their tasks in their social contexts, w.r.t. operational constraints. This happens in contrast to other approaches where agents learn by iteratively interacting with a single opponent from the population \cite{Sugawara2014}\cite{Sen2007}, or by playing repeatedly with randomly chosen neighbours \cite{Airiau2014}. 

Our work goes beyond state of the art methods in resolving congestion problems in any domain, where either a centralized agent learns a global policy, or multiple independent (i.e. non-interacting) Q-learners learn their policies, considering the other agents as part of their environment. Exceptions to this is the method proposed in \cite{Kuyer2008}, where instead of collaborative reinforcement learning methods, the max-plus algorithm has been used, and the method proposed in \cite{Tan2019}, where a model for incorporating multiple deep reinforcement learns is proposed. The multiagent reinforcement learning methods that this paper proposes, can be seen as a continuation of the effort reported in \cite{Vouros2017}: Indeed, we propose multiagent Q-learning methods in which each agent interacts with all its neighbours in the coordination graph, towards computing joint policies to resolve DCB problems. We show through experimentation that these methods are efficient to converging into agents’ joint policies, even when the structure of the coordination graph changes, due to the emergence of new pairs of interacting agents. Also, experiments with collaborative methods provide evidence, in agreement with the results in \cite{Vouros2017}, that agents through collaboration provide solutions of better quality than methods where agents learn in isolation from others.

It is true that the choice of the reward function is critical to the efficiency and effectiveness of the multiagent reinforcement learning method used: Indeed, previous efforts study the use of reward functions (e.g. \cite{Agogino2005}\cite{Crespo2012}\cite{Cruciol2013}\cite{Devlin2014}\cite{Proper2013}) and argue on desirable properties of these functions (i.e. factoredness and learnability), using a specific type of reinforcement learners: Independent learners. Difference rewards have been utilized in multiagent congestion problems \cite{Devlin2014} and have repeatedly demonstrated to help with credit assignment by shaping the global reward, to reward agents contributing to the system's performance and punish agents that do not.  Difference rewards have also been used in the Air Traffic Management domain \cite{Agogino2012}. While these have been shown to satisfy desirable properties when independent learners are used, these are computationally demanding, thus only approximations are used, while their suitability in collaborative settings, is a research aspect that should be thoroughly investigated: This is deemed important in settings where phenomena emerge due to agents’ joined action. In this work we propose a local reward function that is shaped to estimate the cost incurred to an agent due to its participation in hotspots and due to its own delay. The properties of this reward function are thoroughly discussed in section 6.

Finally, and in contrast to the rich literature and various formulations of demand-capacity balance problem in the context of the Air Traffic Flow Management problem (among which those mentioned above) at the tactical phase (i.e. during operation), we consider the DCB problem at the pre-tactical phase.  Thus, conflicts resolved by air traffic controllers (ATC) during the tactical phase are not within the scope of our work. This, in conjunction to considering trajectories as the main objects regulated, supports moving to collaborative decision making for the planning of trajectories, resolving Air Traffic Flow Management problems at the pre-tactical phase, also considering real-world phenomena due to dynamic effects emerging by means of trajectory interactions, and dynamic changes in the airspace sectors’ configurations, including changes in sectors’ capacities.

\section{Multiagent DCB Policy Search Problem Formulation}
According to the problem specification stated above, and using the model of multiagent MDP framework \cite{Guestrin2003}, we formulate the multiagent DCB policy search problem as an MDP comprising the following constituents: 

\begin{itemize}
\item A {\em time horizon} \textbf{H} and a {\em time step} $t=t_0, t_1, t_2, t_3,..., t_{max}$, where $t_{max}-t_0=\mathbf{H}$.
\item The dynamic {\em society} of agents $(\mathbf{A, E^t})$ at time {\em t}, as described above.
\item A set of {\em agent states}: A local state per agent $A_i$ at time {\em t}, comprises state variables that correspond to (a) the delay imposed to the trajectory $T_i$ executed by $A_i$, ranging to $\mathbf{D_i}={0,…,MaxDelay_i}$, and (b) the number of hotspots in which $A_i$ is involved in. Such a local state is denoted by $s^t_i$. The joint state $s^t_{Ag}$ of a set of agents {\em Ag} at time {\em t} is the tuple of all agents in {\em Ag} local states. A {\em global (joint) state} $\mathbf{s}_t$ at time {\em t} is the tuple of all agents' local states. The set of all {\em joint states} for any subset {\em Ag} of agents is denoted $\mathbf{State}_{Ag}$, and the set of {\em joint society states} is denoted by $\mathbf{State}$.
\item The set of {\em agent strategies}: A local strategy for agent $A_i$ at time {\em t}, denoted by $str^t_i$ is the delay for $A_i$ at that specific time point. This delay results from agent’s decisions (actions) at any time point: At each time point until take-off the agent has to take a binary decision. It may either add to its total delay a {\em unit of time}, or not.  The {\em joint strategy} of a subset of agents {\em Ag} of \textbf{A} at time {\em t} (e.g. of $\mathbf{N}^t(Ai)$), is a tuple of local strategies, denoted by $\mathbf{str}^t_{Ag}$ (e.g. $\mathbf{str}^t_{\mathbf{N}^t(A_i)}$).  The joint strategy for all agents \textbf{A} at any time instant {\em t} is denoted $\mathbf{str}_t$. The set of all {\em joint strategies} for any subset {\em Ag} of \textbf{A} is denoted $\mathbf{Strategy}_{Ag}$, and the set of {\em joint society strategies} is denoted by $\mathbf{Strategy}$.
\item The {\em state transition function} {\em Tr} gives the transition to the joint state $\mathbf{s}^{t+1}$ based on the joint strategy $\mathbf{str}^t$ taken in joint state $\mathbf{s}^t$. Formally:
\begin{equation}
Tr: \mathbf{State} \times \mathbf{Strategy} \rightarrow \mathbf{State},
\end{equation}
It must be noticed that the state transition per agent is stochastic, given that no agent has a global view of the society, of the decisions of others, and/or of changing sector configurations, while its {\em neighbourhood} gets updated. Thus, no agent can predict how the joint state can be affected in the next time step. Thus, from the point of view of agent $A_i$ this transition function is actually:
\begin{equation}
Tr: State_{A_i} \times Strategy_{A_i} \times  State_{A_i} \rightarrow [0,1],
\end{equation}
denoting the transition probability $p(s^{t+1}_i|s^t_i , str^t_i)$. 
\item The {\em local reward} of an agent $A_i$, denoted $Rwd_i$, is the reward that the agent gets in a specific state at time {\em t}. The {\em joint reward}, denoted by $Rwd_{Ag}$ for a set of agents {\em Ag} specifies the reward received by agents in {\em Ag} by executing their trajectories according to their {\em joint strategy}, in a {\em joint state}. Further details on the reward function are provided in section 6.
\item A {\em (local) policy} of an agent $A_i$ is a function $ \pi_i: \mathbf{State}_{A_i} \rightarrow \mathbf{Strategy}_{A_i}$ that returns local strategies for any given local state, for agent $A_i$ to execute its trajectory. 
\end{itemize}

The objective for any agent in the society is to find an optimal policy $\pi_i^*$  that maximizes the expected discounted future return: 

\begin{equation}
V_i^*(s) = max_{\pi_i^*}E[\sum_{t=1}^{\infty} \gamma^{t-1}Rwd_i(s_i^t,
\pi_i^*(s_i^t)) | s = s_i^1]
\end{equation}
 
for each state $s_i^t$ for $A_i$, given the initial state $s_i^1$. The discount factor $\gamma$ ranges in [0,1]. 

This model assumes the Markov property, assuming also that rewards and transition probabilities are independent of time. Thus, the state next to state {\em s} given a (joint) strategy is denoted by {\em s'} and it is independent of time. Subsequently, subscripts and superscripts are avoided in cases where it is clear where a state or strategy refers to.

\begin{figure}
\centering
\subfloat[]{\includegraphics[width=0.4\linewidth]{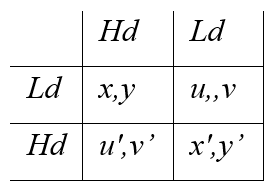}}
\subfloat[]{\includegraphics[width=0.4\linewidth]{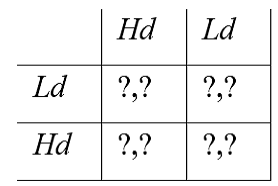}}
\caption{Payoff matrices for 2X2 games.}
\label{fig:payoff}
\end{figure}

The problem is a specific instance of the problem specified in \cite{Vouros2017}, where each agent has several options (minutes of delay in our case) to execute a single task (trajectory in this case), and tasks interact among themselves given some operational constraints. As proposed there, the problem can be formulated as a Markov game: Let us consider a coordination graph with two agents executing interacting trajectories (i.e. trajectories crossing the same sector) and causing a hotspot. Let us also assume that each agent has two delay options: a low-delay (Ld) and a high-delay (Hd). Assuming, without loss of generality, that one of the agents should have a Hd strategy to resolve the DCB problem (otherwise either there would be no hotspot, or the hotspot occurs later in time), agents are assumed to play a game of the form shown in Fig.\ref{fig:payoff}(a): All entries in this matrix are different than zero; {\em x, x, y, y’} can be considered positive integers (in case the hotspot is resolved); {\em u, v, u’, v’} can be negative integers (in case agents do participate in the hotspot). As it can be noticed, this can be a coordination game, with two Nash equilibria, namely the joint options providing payoffs {\em (x, y)} or {\em (x’, y’)}. However, this is not necessarily a symmetric game, considering that the payoff incorporates agents’ preferences and delay costs. The game can be extended to multiple strategies and/or agents executing interacting trajectories. 

Given that the information concerning the effects of agents’ joint decision is not known to any agent in the society, and given that agents do not know about the payoffs of other agents when choosing specific delay strategies, agents need to learn about the structure of the game to be played, and they have to coordinate with others, as well. The information that an agent has about a $2 \times 2$ game is as shown in Fig.\ref{fig:payoff}(b). Question marks indicate the missing information: For instance, none of the two agents knows whether a strategy is effective in resolving hotspots, nor the payoffs from joint strategies.  Our goal is any pair of interacting agents $A_k, A_l$ in the society to converge to a joint delay strategy, so as to resolve hotspots that occur, jointly with all society members.

\section{MARL Algorithms}
We now describe the proposed MARL methods to deal with the multiagent DCB policy search problem. Q-functions, or action-value functions, represent the future discounted reward for a state {\em s} when acting with a specific strategy {\em str} for that state and behaving optimally from then on \cite{Sutton2014}. The optimal policy for any agent $A_i$ in state {\em s} is the one maximizing the expected future discounted reward, i.e. 

\begin{equation}
\pi^*(s)=argmax_{str}Q(s,str).
\end{equation}

In the next paragraphs we describe multiagent reinforcement learning approaches, considering that agents do not know the transition model, and interact concurrently with all their neighbours in the society.

\subsection{Independent Reinforcement Learners (IRL)}
In the Independent Reinforcement Learners (IRL) framework, each agent learns its own policy independently from the others and treats other agents as part of the environment. This is the main paradigm for reinforcement learning agents in the literature for resolving congestion problems.

The independent Q-learning variant proposed in \cite{Guestrin2002} decomposes the {\em global Q-function} into a linear combination of {\em local agent-dependent Q-functions}:
\begin{equation}
Q(\mathbf{s, str}) =\sum_{i=1}^{|A|} Q_i (\mathbf{s}_i,\mathbf{str}_i)
\end{equation}
Each local value, $Q_i$, for agent $A_i$ is calculated according to the local state, $\mathbf{s}_i$, and the local strategy, $\mathbf{str}_i$. The local value $Q_i$ is updated according to the temporal-difference error, as follows:
\begin{eqnarray}
Q_i(\mathbf{s}_i,\mathbf{str}_i) = Q_i(\mathbf{s}_i,\mathbf{str}_i)+ \alpha[Rwd_i(s_i,str_i) + \nonumber \\  
+ \gamma max_{str}Q_i(\mathbf{s}'_i,\mathbf{str}) - Q_i(\mathbf{s}_i,\mathbf{str}_i)]
\end{eqnarray}
It must be noted that instead of the global reward $Rwd(\mathbf{s, str})$ used in \cite{Guestrin2002}, we use the  reward $Rwd_i$ received by the agent $A_i$, taking into account only the local state and local strategy of that agent. Thus, this method is in contrast to the approach of Coordinated Reinforcement Learning model proposed in \cite{Guestrin2002}, since that model needs agents to know the maximizing {\em joint} action in the next state, the associated maximal expected future return, and needs to estimate the Q-value in the global state. 

\subsection{Edge Based Collaborative Reinforcement Learners (Ed-MARL)}

The Edge Based Collaborative Reinforcement Learning (Ed-MARL) is a variant of the Edge Based update sparse cooperative Q-learning method proposed in \cite{Kok2006}. Given two neighbour agents $A_i$ and $A_j$ connected by an edge in the {\em coordination graph} (i.e. representing interacting flights), the {\em Q-function} for these agents is denoted as $Q_{ij}(\mathbf{s}_{ij},\mathbf{str}_{ij})$, where $\mathbf{s}_{ij}$, with abuse of notation, denotes the joint state related to the agents $A_i$ and $A_j$, and $\mathbf{str}_{ij}$ denotes the joint strategy for these two agents. 

Half the sum of all edge-specific Q-functions defines the global Q-function, i.e.
\begin{equation}
Q(\mathbf{s, str}) = \frac{1}{2} \sum_{i,j \in \mathcal{E}} Q_{ij}(\mathbf{s}_{ij},\mathbf{str}_{ij})
\end{equation}
The Q-learning update rule in this case is given by the following equation:
\begin{eqnarray}
Q_{ij}(\mathbf{s}_{ij},\mathbf{str}_{ij}) = (1-\alpha) Q_{ij}(\mathbf{s}_{ij},\mathbf{str}_{ij})+ \nonumber \\ 
+ \alpha \left[ \frac{r_i}{|N(A_i)|} + \frac{r_j}{|N(A_j)|} + 
\gamma Q_{ij}(\mathbf{s}'_{ij},\mathbf{str}^*_{ij}) \right]
\end{eqnarray}
where t is the time instant that the agents reach the joint state $\mathbf{s}_{ij}$, and $\mathbf{str}^*_{ij}$  in \cite{Kok2006} is the best joint strategy for agents $A_i$ and $A_j$ for the joint state $\mathbf{s}'_{ij}$.

In our method, the strategy $\mathbf{str}^*_{ij}$ comprises the best strategy known by agents for the occurring state and it is depicted directly from the agent's value function, $Q_i(s,str)$, which is calculated as the summation of local $Q_{ij}$ values in its neighbourhood:
\begin{equation}
str^* =  argmax_{str}Q_i(s,str)
\end{equation}
\begin{equation}
Q_i(s, str_i) = \sum_{A_j \in N(A_i)}Q_{ij} (\mathbf{s}_{ij},\mathbf{str}_{ij})
\end{equation}
where, $\mathbf{s}_{ij}$ is the agents’ joined state having s as one of its components. This approximates the best strategy in each state, which is improved as the agents learn. We experimentally found out that this approximation method offers comparable quality and considerable improvement in methods’ computational efficiency than using computationally/communication intensive approximation methods, such as the max-sum method used in \cite{Kok2006}. Succinctly, according to the Ed-MARL approach agents update Q-values by propagating edge-specific temporal differences to their neighbouring agents, and only along the corresponding edges, sharing their local rewards with their neighbours.

The main difference between the collaborative Ed-MARL and the IRL method, is that while IRL agents consider own states and strategies without sharing any information with others (i.e. treating others as “noise”), the collaborative approach assumes interacting agents, supporting them to explore joint policies, by means of computing {\em “joint”} Q-values. 

\section{Reward Function}

In many multiagent reinforcement learning problems, the task of determining the reward function in order to produce good performance is quite demanding. The reward function can be approximated in cases where training examples provide ground truth, e.g. via inverse reinforcement learning methods, which is not the case for the DCB problem considered in this work (although it happens in other instances of the problem, as in \cite{Bloem2015}). The utilization of a reward that facilitates coordination is crucial to the majority of multiagent problems, especially to learning in dynamic and complex environments where the actions of individual agents affect the local and global agents’ payoff.

\subsection{Properties of reward functions}

Considering a multiagent system where (a) each agent $A_i$ is taking actions to maximize its local reward $Rwd_i$, and (b) the performance of the full system is measured by the global reward {\em Rwd}; and assuming that the system joint state $\mathbf{s}$ is decomposed into a component that depends on the state of agent $A_i$, denoted by $\mathbf{s}_i$, and a component that does not depend on the state of agent $A_i$, denoted by $\mathbf{s_{-i}}$, we aim to produce rewards that facilitate learning joint strategies efficiently. In doing so, we may consider two properties of the reward function, already proposed in \cite{Agogino2005}: 

The first property, called {\em factoredness}, concerns “aligning” the individual agent rewards with the global reward. For an agent $A_i$, the degree of factoredness between the rewards $Rwd_i$, and {\em Rwd}, at joint state $\mathbf{s}$ is defined as:
\begin{equation}
F_{Rwd_i}=\frac{(\sum_{s'}u[(Rwd_i(s)-Rwd_i(s'))(Rwd(s)-Rwd(s'))])}{(\sum_{s'} 1)}
\end{equation}
where the states {\em s’} and {\em s} only differ in the states of agent $A_i$, and $u[x]$ is the unit step function, equal to 1 if $x > 0$. Intuitively, the degree of factoredness gives the percentage of states in which the action of agent $A_i$ has the same impact on $Rwd_i$ and {\em Rwd}. 

The second property, called {\em learnability}, measures the dependence of the reward on the actions of a particular agent as opposed to all the other agents. The {\em point learnability} of reward $Rwd_i$ between state {\em s} and {\em s’} is defined as the ratio of the change in $Rwd_i$ due to a change in the state of agent $A_i$ over the change in $Rwd_i$ due to a change in the states of other agents:
\begin{equation}
L(Rwd_i,s,s')= \frac{\parallel Rwd_i(s)- Rwd_i(s-s_i+s'_i)\parallel}{\parallel Rwd_i(s)-Rwd_i(s'-s'_i+s_i )\parallel}
\end{equation}

where, addition and subtraction operators remove or add components into states.

The learnability of a reward $Rwd_i$ is then given by:
\begin{equation}
L(Rwd_i,s)=\frac{\sum_{s'}L(Rwd_i,s,s')}{(\sum_{s'} 1)}
\end{equation}
Intuitively, the higher the learnability, the more $Rwd_i$ depends on the move of agent $A_i$, i.e., the better the associated signal-to-noise ratio for $A_i$.

\subsection{The proposed reward function}

For the DCB problem we have formulated the individual delay reward $Rwd_i$: For an agent $A_i$ in \textbf{A} this reward depends on the participation (contribution) of that agent in hotspots occurring while executing its trajectory, according to its strategy $str^t_i$ in state $s^t_i$, i.e. according to its decided delay. Formally:
\begin{equation}
Rwd_i(s^t_i, str^t_i) = C(s^t_i, str^t_i) - \lambda * DC(str^t_i)
\end{equation}
where, 
\begin{itemize}
\item $C(s^t_i, str^t_i)$ is a cost function that depends on the participation of $A_i$ in hotspots, given the strategy $str^t_i$, and
\item $DC(str^t_i)$ is a function of agent’s ground delay cost given by that strategy. 
\end{itemize}
The (user-defined) parameter $\lambda$ balances between the cost the participation in hotspots implies, and the cost of the ground delay imposed towards. The goal is for any agent to participate in zero hotspots with the minimum possible delay. 

Actually, both functions $C(s^t_i, str^t_i)$ and $DC(str^t_i)$ represent delay costs at the pre-tactical phase of operations. We have chosen $C(s^t_i, str^t_i)$ to depend on the total duration of the period in which  agents fly over congested sectors: This gives a measure of the severity of the imbalances in which agent $A_i$ contributes, and it is equal to the maximum delay that it may get if it is the sole agent causing the congestion. This is multiplied by 81 which is the average strategic delay cost per minute (in Euros) in Europe when 92\% of the flights do not have delays \cite{Cook2015}. If there is not any congestion, then this is a positive constant that represents the reward agents get by not participating in any hotspot. Overall, the actual form of $C(s^t_i, str^t_i)$ is as follows:
\begin{equation}
C(s^t_i, str^t_i) = \left\{ \begin{array}{ccl}
    -TDC \times 81 &, & \mbox{if } TDC>0 \\ 
    PositiveReward &, & \mbox{if } TDC = 0 
\end{array}\right.
\end{equation}
where, {\em TDC} is the total duration for agent $A_i$ in congested sectors. The first case holds when there are hotspots in which the agent participates (thus, the total duration in hotspots, TDC, is greater than 0), while the second case holds when agents do not participate in hotspots.

The $DC(str^t_i)$ component of the reward function corresponds to the {\em strategic delay cost} when flights delay at gate. In our implementation, this depends solely on the minutes of delay and the aircraft type, as specified in \cite{Cook2015}. As such, the actual form of this function is as follows:
\begin{equation}
DC(str^t_i) = StrategicDelayCost(str_i^t,Aircraft(A_i))
\end{equation}
where $str^t_i$ is the delay imposed to the agent $A_i$ and {\em StrategicDelayCost} is a function that returns the strategic delay cost given the aircraft type of agent $A_i$ and its delay. In the general case the function $DC(str^t_i)$ could be taking into account additional airline-specific strategic policies and considerations regarding flight delays.

Coming to the properties of the individual delay reward function, and specifically to its factoredness, the impact $Rwd_i$ has on the global reward {\em Rwd}, given as the summation of individual agents’ rewards, depends on the effect that the strategy of $A_i$ has on the states of other agents. Specifically, we may distinguish the following cases: (a) For agents that do not interact with $A_i$, i.e. for agents with which $A_i$ does not co-occur in hotspots, the action of $A_i$ has not any {\em direct} effect to their reward (it may affect their reward, but only indirectly via the strategy of other agents). Specifically, for these agents, neither the {\em C} nor the {\em DC} part of their reward changes due to the action of $A_i$. (b) For agents in $N(A_i)-A_i$ with whom $A_i$ interacts, and in case these agents do not change their strategy, their reward changes only by changing $A_i$'s strategy. In this case, the {\em C} part of their reward may change due to a possible change in occurring hotspots, and the {\em DC} part changes due to the change of $A_i$ delay, only. (c) For agents in $N(A_i)-A_i$ with whom $A_i$ interacts, and in case these agents do change their strategy synchronously to $A_i$, then the strategy of $A_i$ impacts the rewards of these agents, given that rewards consider the hotspots occurring due to agents’ joint action ({\em C} part of the reward). On the other hand, while each agent considers the cost incurred due to its own strategy ({\em DC} part of the reward) only, the total reward is aligned with individual agents’ rewards.

As a conclusion, the individual delay reward has a high degree of factoredness given that rewards of agents in the society are affected by the strategies of others, and the global reward is aligned with individual agents’ rewards. 
As far as learnability is concerned, in the DCB domain, as in other domains where the joint strategy of agents is of importance, the reward received by the agent $A_i$ depends on the joint strategy of all agents: The resulting hotspots emerge as an effect of agents’ joint strategy, even if agents did not interact directly with $A_i$. Thus, the ratio of the change in $Rwd_i$ due to a change in the state of agent $A_i$, over the change in $Rwd_i$ due to a change in the states of other agents, may not be always proportional to the effectiveness of agent’s strategy, as it depends on the strategies of the other agents.

\section{Experimental Results}

\subsection{Description of evaluation cases}

To evaluate the proposed MARL methods, we have constructed evaluation cases of varying difficulty. Appendix A describes the general procedure we follow for constructing an evaluation case. Although the difficulty of DCB problems cannot be determined in a rigorous way, we did this empirically, by inspecting problem parameters (explained subsequently) in conjunction to the average delay per flight according to the delays imposed by the Network Management organization (NM). While the NM specifies the delay to be imposed to each flight towards resolving demand-capacity imbalances, this is not always a DCB problem solution: Hotspots do occur even if delays are imposed to flights. This shows the tolerance of the system, as well its reliance to resolving imbalances in the tactical phase of operations, as opposed to resolving hotspots in the pre-tactical phase, according to our aim. Having said that, it is important to point out that delays imposed by the NM cannot be compared in a direct way to solutions provided by the proposed methods, given that in the former case many decisions are to be taken at the tactical phase. However, comparison shows the merits of MARL methods to solve DCB problems.

\begin{table*}
\begin{center}
\caption{Description of the evaluation cases used in our experimental study.}
 \begin{tabular}{|c|c|c|c|c|c|c|} \hline
{\bf Evaluation} & \textbf{\emph{\# flights}} & \textbf{\emph{Average Degree in}} & \textbf{\emph{MaxDelay}} &  \textbf{ \emph{Average}} & \textbf{\emph{\# flights}} & \textbf{\emph{\# hotspots}} \\ 
{\bf Case ID} & & \textbf{\emph{Coordination Graph}} & &  
\textbf{ \emph{Delay}} & \textbf{\emph{with}} & 
\textbf{\emph{(participating)}} \\
& & \textbf{\emph{(non-zero min/max)}} & & & 
\textbf{\emph{delay}} & 
\textbf{\emph{(flights)}} \\ \hline

Aug4 &	5544 &	 6.41(17-20) &	66  & 0.383 &	146 &   33 (853)\\ \hline
Aug13 &	6000 &	10.89(22-105) &	147 & 1.152 &	415 &	53 (1460)\\ \hline
Jul2 &	5572 &	6.39(29-107)  &	80  & 1.663 &	498 &	29 (778)\\ \hline
Jul12 &	5408 &	5.84(21-95)  &	95  & 0.95  &	254 &	28 (820)\\ \hline
Sep3 &	5788 &	5.24(18-77)  &	61  & 0.732 &	280 &	26 (783)\\ \hline
\end{tabular}
\label{EvaluationCases}
\end{center}
\end{table*}

Each evaluation case corresponds to a specific day in 2016, above Spain, and its {\em “difficulty”} has been determined by means of the {\em number of flights involved}, the {\em average number of interacting flights per flight} (which is translated to the average degree for each agent in the coordination graph, at starting time $t_0$), the {\em maximum delay} imposed to flights for that day to resolve DCB problems according to the NM, the {\em average delay} per flight for all flights according to NM, and the {\em number of hotspots} in relation to the number of flights participating in these hotspots, at starting time $t_0$.

Table \ref{EvaluationCases} presents the different cases, named by the day to which they correspond. Columns specify the following:

\begin{itemize}
\item \textit{Number of flights:} The number of flights for that particular day above Spain.
\item \textit{Average Degree in Coordination Graph (min/max):} This indicates in average the traffic (i.e. the number of interacting flights) experienced by each of the agents (flights) in each evaluation case at starting time $t_0$. This number changes while agents decide on their strategies. It is expected that as the coordination graph becomes denser, i.e. as the average degree of nodes in the coordination graph increases, the problem becomes more computationally demanding. Min/Max indicates the minimum and the maximum degree reported in the coordination graph per evaluation case at starting time $t_0$, ignoring zeros.
\item \textit{MaxDelay:} This is the maximum delay imposed to any flight according to the NM for that particular day, and we use this in our experiments as the maximum delay that can be imposed to any flight for the corresponding case.
\item \textit{Average Delay:} This is the average delay per flight reported by the NM for that particular day, ignoring all delays with less than 4 min (considered as “no delay”, according to experts advise).
\item \textit{Number of flights with delay:} This is the number of flights with delays more than 4 min, due to demand-capacity imbalances, as reported by NM for that particular day. Delays due to other reasons (e.g. weather conditions) are ignored.
\item \textit{Max number of hotspots (number of flights):} It indicates the number of hotspots of each individual case at starting time $t_0$, together (in parenthesis) with the number of flights that participate in these hotspots (each flight may participate in multiple hotspots).
\end{itemize}

As far as the difficulty of cases is concerned, it turns out that there are many other crucial features that determine the difficulty of each case, such as the duration of flights in hotspots, the specific excess on capacity for these hotspots, the number of consecutive counting periods in which the demand exceeds the capacity of a sector etc. We have not determined the exact features (this is out of scope of this work), but after extensive experimentation, we can conclude that the degree of difficulty is proportional to the average delay per flight that should be imposed, and to the ratio of flights with delay to the flights participating in hotspots at time $t_0$. Specifically, it can be assessed as follows:

\begin{equation}
  \begin{aligned}[b]
      & DegreeOfDifficulty = \\
      & \frac{AverageDelayPerFlight *NumberOfFlightsWithDelay}{FlightsPartcipatingInHotspots}
  \end{aligned}
\end{equation}

\begin{figure}
\centering
\includegraphics[width=0.46\linewidth]{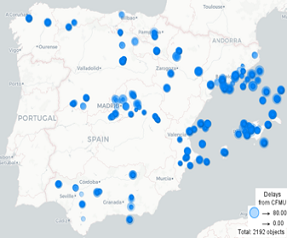} 
\includegraphics[width=0.46\linewidth]{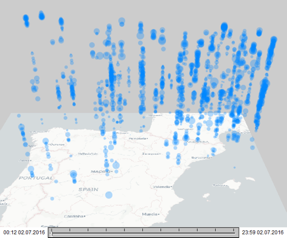}
\caption{NM solution for July 2: (Left) Flight delays are represented by circles positioned at the sector centroids. The sizes are proportional to the delay durations. (Right) The space-time cubes show the spatio-temporal distribution of the delays. The time axis is oriented upwards, towards the end of the day.}
\label{fig:NM_solution}
\end{figure}

Based on this rule the most difficult case among our cases in Table \ref{EvaluationCases} is the case Jul2. To understand the difficulty of the Jul2 case we need to delve into the visualizations shown in Fig.\ref{fig:NM_solution}. Visual exploration of to DCB cases and solutions, provide insights about the situations occurring and the rationale behind imposed delays. We have used the V-Analytics tool \cite{Andrienko2013} to produce maps and space-time cubes that show the spatio-temporal distributions of the delays. Delays are represented by circles and the sizes are proportional to the delay duration. The spatial positions of the cycles are the positions of the sector centroids, depicting delays per sector. The temporal axis in the cubes goes from the bottom to the top.  

The maps show that the NM imposes increased delays on the east (areas of Barcelona, Canary Islands, and Valencia) and on the south (Seville and Granada), while delays on the northwest of Spain, compared to the others, are not that many or large. Space-time cubes show that delays are distributed during the day, except in the northwest of Spain where they increase slightly by the end of the day.

\begin{figure}
\centering
\includegraphics[width=0.5\linewidth]{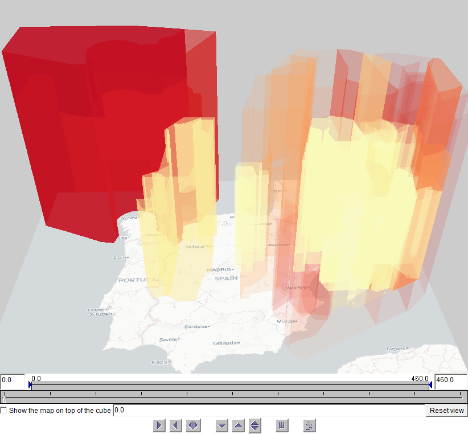}
\includegraphics[width=0.9\linewidth]{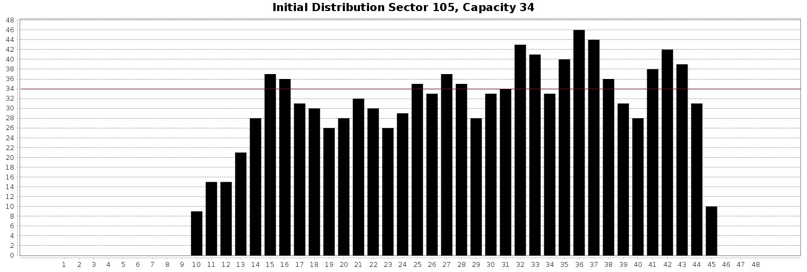}
\caption{(upper) A 3D view shows the 3D shapes of the sectors whose capacities were exceeded with the NM solution provided for Jul2.  The colouring from yellow to red represents the maximal capacity excess. (lower) Demand evolution for the most demanded sector in northwest Spain for Jul2, prior to applying any delay. The capacity of this highly-demanded sector is 34 (indicated with the horizontal red line).}
\label{fig:capacities}
\end{figure}

Concerning the details of the problem, we have extracted the sectors’ capacity excess events from the time series of the sector demand differences to the capacities. Fig.\ref{fig:capacities} shows the 3D shapes of the sectors whose capacities were exceeded with the NM solution provided. The red colour of the north-west sector shows the large demand compared to the capacity for this sector. This is further shown in the histogram in Fig.\ref{fig:capacities} The histogram shows the demand evolution per counting period for the highly-congested sector in the initial problem state (i.e. prior to imposing any delay).

Therefore, while there is an excess of capacity for many sectors, this happens at a large degree in the northwest part of Spain for Jul2: As the evolution of demand shows for this highly-demanded sector (Fig.\ref{fig:capacities}), the excess of capacity is high, especially in consecutive periods at the second-half of the day. This means that while delays may resolve demand-capacity imbalances for specific periods, imbalances in adjacent periods may become more severe.

\subsection{Evaluation Results}

In our experimental study we have evaluated our methods using the following metrics:

\begin{itemize}
\item \textit{Number of flights with delay:} also mentioned as “regulated flights”.
\item \textit{Average delay per flight:} the ratio of total delay to the number of flights ignoring delays less than 4 minutes.
\item \textit{Learning curves:} the progress of the average delay per flight as agents learn. As algorithms converge to solutions, the number of hotspots should be reduced and eventually reach to zero, signifying the computation of a solution, while the average delay should be reduced. Therefore, the speed of reaching that point (zero hotspots) and the round at which methods stabilize\footnote{I.e. remaining to zero hotspots and to a specific value for flights’ average delay -without oscillating between non-solutions and/or solutions, and/or different average delay values.} the agents’ joint policy, signify the computational efficiency of the method to reach solutions. It must be noted that, in case that a method cannot reach a solution for a specific case, it may converge to a joint policy that do not resolve all hotspots.
\item \textit{Distribution of delays to flights:} histograms showing the number of flights in discrete time intervals, such as 5-9 minutes of delay, 10-29 minutes of delay, 30-59 minutes of delay, etc., up to the $MaxDelay$ specified per case.
\end{itemize}

Results reported are averages of results computed by 20 independent experiments per case and method.  Specifically, we report results from two methods: Independent (IRL), and Collaborative (Ed-MARL) Reinforcement Learning. Finally, and in order to show the tolerance of the methods in specific delay preferences per flight, we explore their potential to resolve DCB problems by considering constraints on the $MaxDelay$ imposed to specific flights. In other words, these are cases where some of the flights impose a constraint to their maximum delay, much less than the $MaxDelay$ imposed to the other flights (Appendix B).

The proposed MARL methods have been executed for 15000 rounds (episodes) following an $\epsilon$-greedy exploration-exploitation strategy starting from probability 0.9, which every 120 rounds is diminished by the value of 0.01. At episode 10800 the probability becomes 0.001 and is henceforth considered zero. Then, a pure exploitation phase starts. In addition to the above, and in order to enhance the efficiency of the proposed methods, we have considered a deterministic rule for the flights that do not participate in any hotspot (i.e. agents with no neighbours): These are set to have delay equal to 0. It must be pointed out that any of these flights may participate in hotspots in any past/future state, due to the joint strategies of the other agents. In any such case the corresponding agents participate in the multiagent RL process. 

\begin{figure*}
\centering
\includegraphics[width=0.4\linewidth]{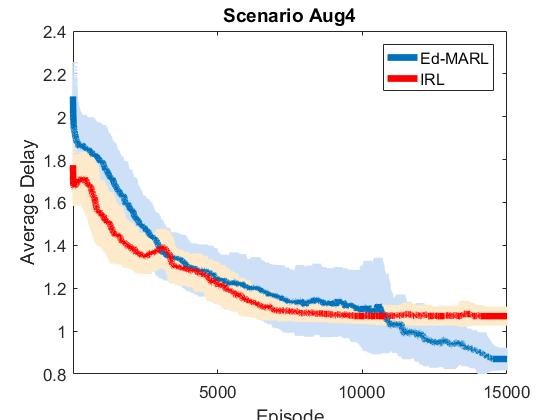} 
\includegraphics[width=0.4\linewidth]{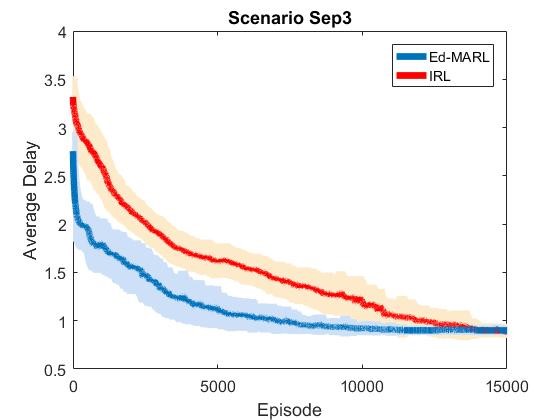}
\includegraphics[width=0.4\linewidth]{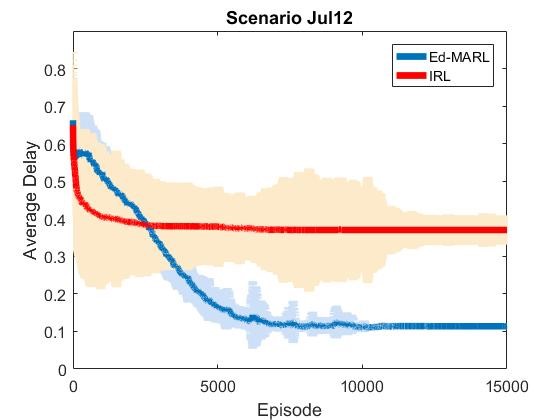} 
\includegraphics[width=0.4\linewidth]{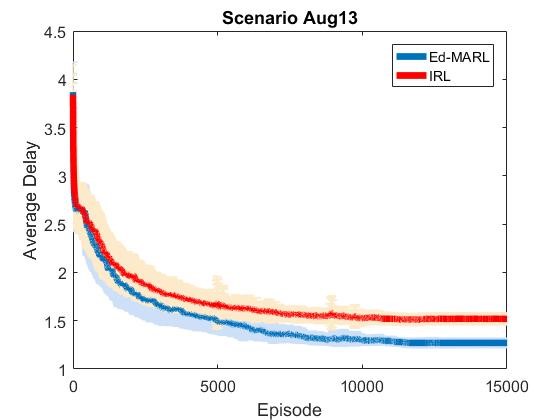}
\includegraphics[width=0.4\linewidth]{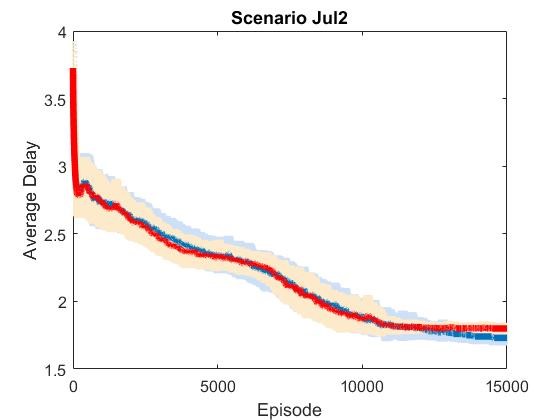} 
\caption{The learning curves of the proposed methods per evaluation case, showing how agents manage to learn joint policies to resolve DCB problems, resulting to 0 hotspots, while reducing the average delay per flight. The x-axis corresponds to the learning episodes, while the y-axis to the average delay per flight.}
\label{fig:learning_curves}
\end{figure*}

Fig.\ref{fig:learning_curves} illustrates the learning curves obtained by the training procedure in terms of the average delay, where we show the mean delay and its standard deviation at every episode of the learning process.  Both methods converge effectively after a period of exploration. In most case their performance is remarkable since they reach (almost) always the same solution with a low variability. However, the Ed-MARL approach offers more qualitative solutions with less average delay per flight. The IRL method manages to converge much faster (earlier than episode 10000) since it is simpler, without any overhead due to the collaboration among agents. However, it should be noticed that “convergence” does not imply solving the problem. A method may converge to a joint policy, imposing delays to flights, which may not resolve all hotspots. This is not the case for the methods, which resolve all imbalances, in all evaluation cases. 

\begin{figure*}
\centering
\includegraphics[width=0.45\linewidth]{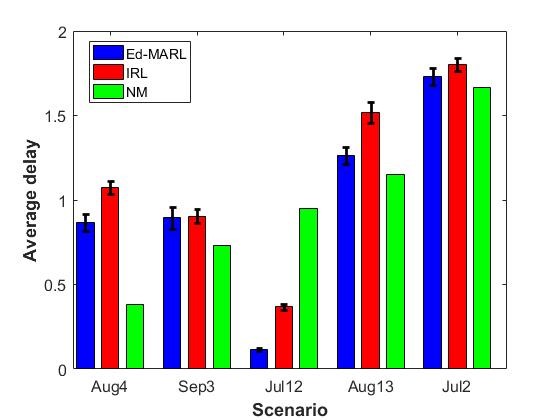} 
\includegraphics[width=0.45\linewidth]{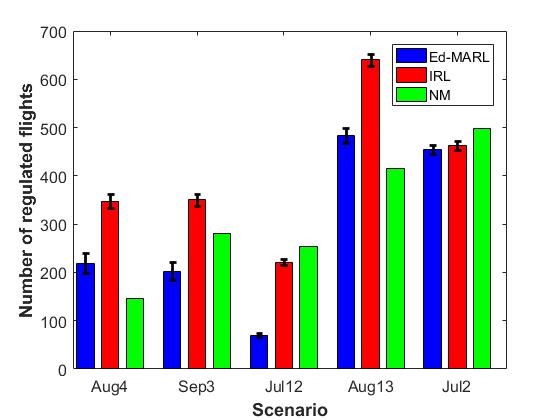}
\caption{Comparative results in bar charts: (a) Average delays per flight and (b) Number of flights with delays from: Ed-MARL (blue bar), IRL (red bar), NM (green bar). The x-axis shows evaluation cases that have been sorted according to NM solutions’ average delay per flight. }
\label{fig:comparative_results}
\end{figure*}

Delving into the quality of results, Fig.\ref{fig:comparative_results} shows the average delay per flight (Fig.\ref{fig:comparative_results}(a)) and the number of regulated flights (Fig.\ref{fig:comparative_results}(b)) reported by each method. In particular, Fig.\ref{fig:comparative_results}(a) provides the average delay (y-axis) per flight on every evaluation case (x-axis). For comparative purposes, we show also the average delay reported by the NM. Evaluation cases are ordered according to the average delay per flight reported by NM in increasing order (green bar).

As it can be observed, the average delay per flight reported by MARL methods does not increase consistently to the average delay per flight reported by NM. This difference with the NM reflects the shift of paradigm MARL methods provide. While the NM imposes delays to flights in a “first come – first regulated” basis, MARL methods regulate flights jointly, so as to reach a solution that is of best interest to all agents. Among the MARL methods, the Ed-MARL (blue bar) is consistently better than the IRL method except for the Sep3 case where both methods achieve similar results in terms of average delay. We can also notice that the NM solution provides lower average delay per flight in every evaluation case except for Jul12. However, as already pointed out, delays imposed by the NM do not resolve all the demand-capacity imbalances. For instance, the NM delays resolve only 2 hotspot occurrences out of 33 in Aug4 scenario. Regarding the number of flights to which our methods impose delays (regulated), as shown in Fig.\ref{fig:comparative_results}(b), the Ed-MARL method manages to provide solutions with consistently less regulated flights than the IRL approach. Again, the number of regulated flights reported by the NM is lower than the flights regulated by the proposed methods (except for the Jul2 and Jul12 cases), but they do not manage to resolve all the imbalances. Looking at the results, we can conclude that the Ed-MARL collaborative method provides the most promising results. Visualizations of their solutions provide more evidence on their quality and on the rationale behind agents’ strategies.

\begin{figure*}
\centering
\includegraphics[width=0.45\linewidth]{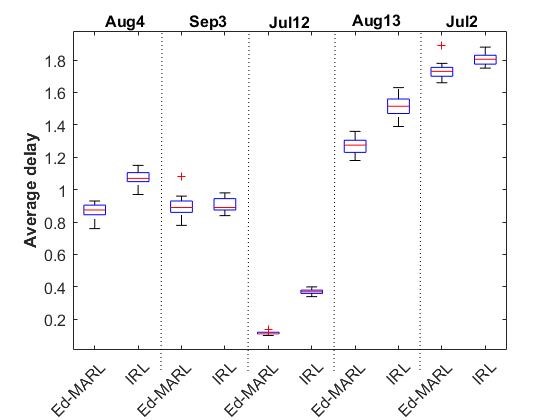} 
\includegraphics[width=0.45\linewidth]{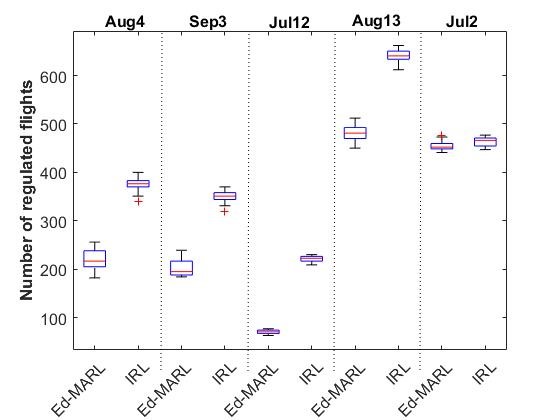}
\caption{Comparative results of (a) average delays and (b) number of regulated flights, presented in box plots for Ed-MARL and IRL methods per evaluation case.}
\label{fig:box_plots}
\end{figure*}

\begin{table*}[]
\caption{Statistical measurements of average delays obtained by 20 intependent experiments.}
\begin{tabular}{|c|c|c|c|c|c|c|c|c|}
\hline
\multirow{2}{*}{\textbf{Evaluation Case ID}} & \multicolumn{4}{c|}{\textbf{Ed-MARL}}                            & \multicolumn{4}{c|}{\textbf{IRL}}                                \\ \cline{2-9} 
                                    & \textbf{Avg} & \textbf{Std} & \textbf{Median} & \textbf{P-Value} & \textbf{Avg} & \textbf{Std} & \textbf{Median} & \textbf{P-Value} \\ \hline
\textbf{Aug4} &0.868    &0.049  &0.875  &0.443  &1.074  &0.042  &1.070           &0.691  \\ \hline
\textbf{Sep3} &0.895	&0.063	&0.890	&0.794	&0.905	&0.042	&0.890	&0.222   \\ \hline
\textbf{Jul12} &0.112	&0.006	&0.110	&0.018	&0.369	&0.017	&0.370	&0.726\\ \hline
\textbf{Aug13} &1.266	&0.049	&1.275	&0.952	&1.516	&0.062	&1.515	&0.996 \\ \hline
\textbf{Jul2} &1.729	&0.050	&1.730	&0.556	&1.802	&0.037	&1.805	&0.480 \\ \hline
\end{tabular}
\label{Stats_AvgDelay}
\end{table*}

\begin{table*}[]
\caption{Statistical measurements of number of regulated flights obtained by 20 intependent experiments.}
\begin{tabular}{|c|c|c|c|c|c|c|c|c|}
\hline
\multirow{2}{*}{\textbf{Evaluation Case ID}} & \multicolumn{4}{c|}{\textbf{Ed-MARL}}                            & \multicolumn{4}{c|}{\textbf{IRL}}                                \\ \cline{2-9} 
                                    & \textbf{Avg} & \textbf{Std} & \textbf{Median} & \textbf{P-Value} & \textbf{Avg} & \textbf{Std} & \textbf{Median} & \textbf{P-Value} \\ \hline
\textbf{Aug4} &219.4	&20.1	&216.5	&0.723	&374.7	&14.3	&376.5	&0.722  \\ \hline
\textbf{Sep3} &202.7	&18.2	&195.0	&0.275	&350.0	&12.4	&351.0	&0.582   \\ \hline
\textbf{Jul12} &70.42	&4.0	&71.0	&0.739	&221.1	&6.3	&223.0	&0.787\\ \hline
\textbf{Aug13} &482.1	&15.8	&481.0	&1.000	&639.9	&13.1	&640.5	&0.963 \\ \hline
\textbf{Jul2} &454.2	&9.8	&452.0	&0.694	&463.4	&9.4	&466.0	&0.645 \\ \hline
\end{tabular}
\label{Stats_Flights}
\end{table*}

We further present in Fig.\ref{fig:box_plots} the results of both MARL methods using box plots as calculated by executing 20 independent experiments at every evaluation case. The left diagram Fig.\ref{fig:box_plots}(a) shows the {\em average delay} per flight and the right diagram Fig.\ref{fig:box_plots}(b) the {\em number of regulated flights}. Moreover, in an attempt to gain a clearer picture of the performance of our methods, we provide in Tables \ref{Stats_AvgDelay}, \ref{Stats_Flights} several statistical measures calculated by 20 experiments: {\em mean value (Avg)}, {\em standard deviation (Std)}, {\em median} and {\em the p-value} of test analysis, respectively. The last statistic, p-value, refers to the {\em Kolmogorov-Smirnov (K-S) test} that performs a test for normal distribution of data. The greater this value, the most evidence to accept the null hypothesis, i.e. that data follow a normal distribution, and to consider the obtained experimental results more statistically significant and representative. From these results we can see the small variability of two evaluation measurements, as well as the high p-values. Only in one case (Jul12) the obtained average delays of the Ed-MARL method gave a small p-value (0.016). However, looking carefully in Table \ref{Stats_AvgDelay}, its standard deviation (Std) is very small (0.009) that indicates almost constant behavior.

Results can be further explained by the conjunction of the following facts: 
\begin{itemize}
\item The collaborative Ed-MARL method guide agents to form joint policies within their neighbours by sharing their local rewards via edges in the coordination graph.
\item Any agent using the IRL method does not have the capability to shape joint strategies with peers via sharing rewards, and does not affect in any direct way the Q-values learnt by other agents.
\end{itemize}

\begin{figure*}
\centering
\includegraphics[width=0.45\linewidth]{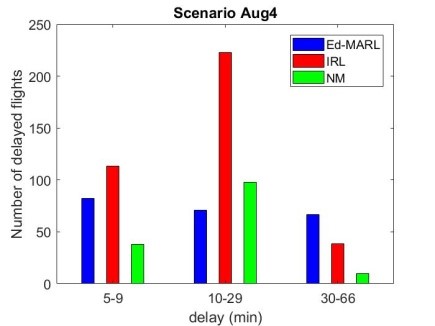} 
\includegraphics[width=0.45\linewidth]{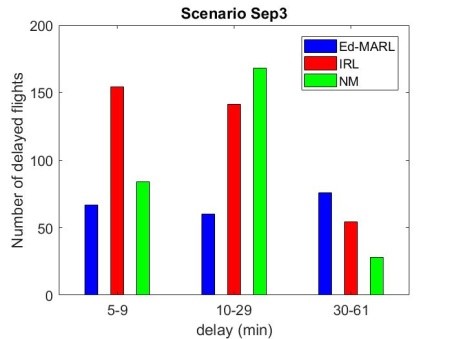}
\includegraphics[width=0.45\linewidth]{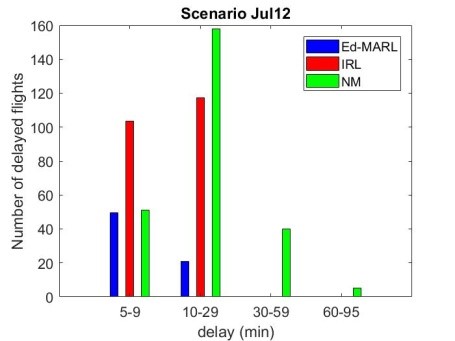} 
\includegraphics[width=0.45\linewidth]{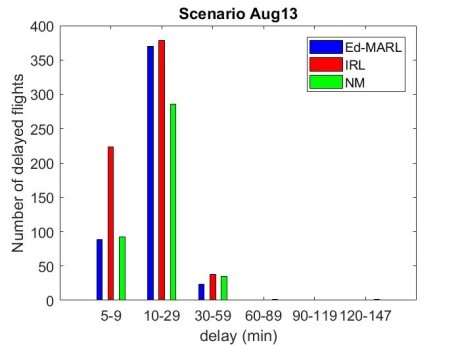}
\includegraphics[width=0.45\linewidth]{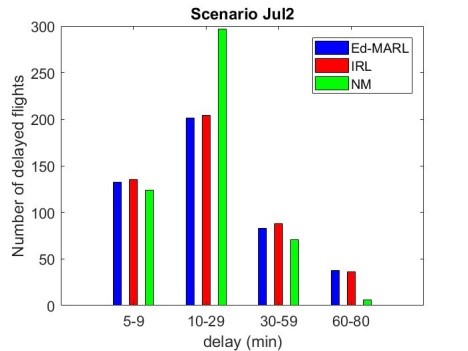} 
\caption{The distribution of delays to flights per evaluation case. The x-axis indicates the delay imposed while the y-axis corresponds to the number of flights with delay. Notice that the maximum delay (and thus x-axis values) differs among evaluation cases.}
\label{fig:delay_distribution}
\end{figure*}

Furthermore, Fig.\ref{fig:delay_distribution} provides the distribution of delays to flights reported by both MARL methods and the NM, in all evaluation cases. Although it seems that the NM avoids imposing large delays on flights, we need to recall that the NM delays do not provide solutions to the DCB problems. The Ed-MARL method ends up in giving delays to fewer flights, while both methods reduce drastically the number of regulated flights as moving from small to large delay intervals. A last remark is about the increased number of flights with large delays in three of the cases (Aug4, Sep3, Jul2). Using appropriate visualizations that show distribution of delays in space and time we can delve into the details of each case and offer further advise to choosing solutions, explaining the rationale behind methods’ solutions. We discuss the solution for the most difficult case: Jul2.

\begin{figure}
\centering
\includegraphics[width=0.46\linewidth]{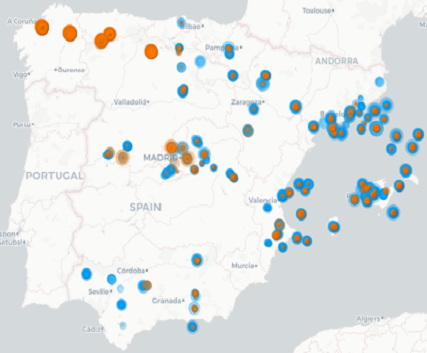} 
\includegraphics[width=0.46\linewidth]{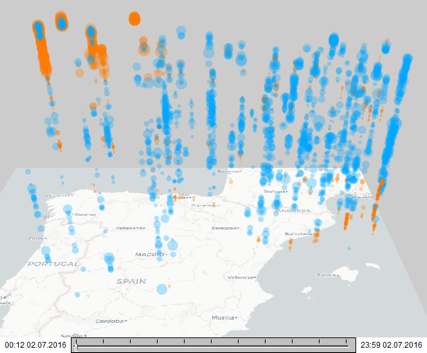}
\caption{Ed-MARL solutions (in orange) for Jul2 compared to NM solutions (in blue).}
\label{fig:ed_marl_solution}
\end{figure}

\begin{figure}
\centering
\includegraphics[width=0.46\linewidth]{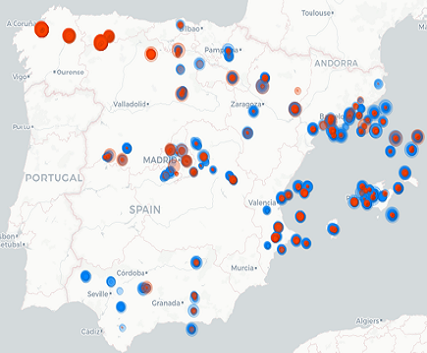} 
\includegraphics[width=0.46\linewidth]{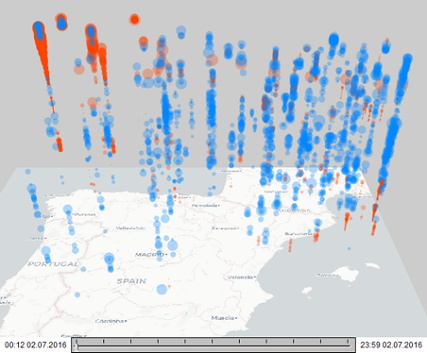}
\caption{IRL solutions (in orange) for Jul2 compared to NM solutions  (in blue).}
\label{fig:irl_solution}
\end{figure}

Figs. \ref{fig:ed_marl_solution} and \ref{fig:irl_solution} present spatial and spatio-temporal distribution of delays imposed by MARL methods, superimposed to NM delays. In particular, Fig.\ref{fig:ed_marl_solution} presents the solution from the Ed-MARL method, while Fig.\ref{fig:irl_solution} presents the solution from the IRL method. The maps show that, compared to the NM, the methods reduce the delays on the east (areas of Barcelona, Canary Islands, and Valencia) and on the south (Seville and Granada) but increase the delays on the northwest of Spain. Space-time cubes also show that MARL methods perform notably better in the first half of the day. In all areas except the northwest of Spain they also perform well in the second half of the day. The delays in the northwest area significantly increase by the end of the day according to both MARL methods.

We need to recall that extracting the sectors’ capacity excess events from the time series of the sector demand, we observed an excess of capacity for many sectors, which happens at a large degree in the northwest part of Spain for Jul2, especially in consecutive periods at the second-half of the day. MARL methods, thus, to resolve these hotspots, increase the delays imposed to flights, especially for flights crossing the parts of the airspace in the periods of high demand.

\section{Conclusions}

In this paper we formulate the problem of resolving demand-capacity imbalances (DCB) in ATM as a coordination problem of agents that operate to a multiagent MDP framework. We propose the use of MARL techniques to solve this problem, as a new paradigm for resolving demand-capacity imbalances at the pre-tactical phase of operations. A novel, generic reward function was constructed that takes into account the agents' participation in hotspots, and also the strategic cost of delay. The proposed methods were evaluated on real-world cases encompassing thousands of agents in complex / dynamic settings, where we measure their efficiency (speed of convergence) and effectiveness (quality of solutions). The results were very promising as our collaborative MARL approaches were able to successfully resolve complex DCB problems in ATM.

More than that, we envisage the work laid out in this paper to be seen as a first step towards devising multiagent methods for prescribing the effect of traffic to correlated aircraft trajectories, contributing to the transition to a trajectory-based air-traffic management paradigm. This will hopefully help overcome the shortcomings of the currently used ATM paradigm, and as such, could in time allow commercial aircrafts “to fly their preferred trajectories without being constrained by airspace configurations”.

It is our intention to further pursue and develop the proposed scheme, mainly on the following directions: 

\begin{itemize}
\item Extend our work in deep reinforcement learning schemes by considering continues state spaces and value function approximation models.
\item Validate the proposed framework on other real-world evaluation cases in a more systematically way.
\item Study the effectiveness alternative reward functions that may contain richer discriminating features.
\end{itemize}
\vspace{1cm}
\textbf{Acknowledgment}
This work has been supported by the DART project, which has received funding from the SESAR Joint Undertaking under grant agreement No 699299 under European Union Horizon 2020 research and innovation programme; It has been partially funded by National Matching Funds 2017-2018 of the Greek Government, and more specifically by the General Secretariat for Research and Technology (GSRT), related to DART and datAcron project. For more details, please see the DART project's website, http://www.dart-research.eu.

\section*{References}
\bibliographystyle{elsarticle-harv}
\bibliography{ESA2019}

\newpage

\appendix
\section{Construction of evaluation cases}
The following procedure describes how we have created evaluation cases:

The first step is to collect all planned flight trajectories (Flight Plans) for that day as provided by the Spanish Operational Data. According to the domain experts, we construct evaluation cases using the Flight Plans specified just before take-off. This makes solutions provided by MARL methods comparable to the delays imposed by the NM. Some Flight Plans span in two consecutive days. These Flight Plans are considered to be part of the problem for both days. Finally, all flights participating to the evaluation case are distinguished between commercial and non-commercial. Delays cannot be imposed to non-commercial flights (e.g. military). In addition, the model of each aircraft executing a trajectory is stored for the calculation of strategic delay costs.

After identifying the Flight Plans, we cross-check them with the flights considered by the NM. In doing so, Flight Plans that do not correspond to an NM entry are dropped, and delays imposed by the NM to resolve hotspots occurring inside the Spanish Airspace are identified.

While Flight Plans specify trajectories crossing air volumes, this sequence is exploited to compute the series of active sectors that each flight crosses - depending on the open sector configurations in different periods during the day - together with the entry and exit time per crossed sector. For the first (last) sector of the flight, where the departure (resp. arrival) airport resides, the entry (resp. exit) time is the departure (resp. arrival) time. However, there may exist flights that cross the airspace but do not depart and/or arrive in any of the sectors of our airspace: In that case we consider the entry and exit time from sectors within the airspace of our interest.
Since airspace sectorization changes frequently during the day, we need to exploit the mappings from air volumes to open sectors, “translating” air volumes crossed by trajectories to open sectors. It must be noticed that:

\begin{itemize}
\item Given any delay imposed to a trajectory, sectors crossed may vary due to the changing sector configurations.
\item This may result in alternative representations of a single trajectory; one for each possible delay (each crossing a different set of open sectors).  
\end{itemize}

The trajectories specifying the Flight Plans, in conjunction to the list of all the necessary sectors with their capacities, comprise an {\em evaluation case}.

In addition to the above, each evaluation case contains the following parameters:

\begin{itemize}
\item The \textit{number of flights} (i.e. participating agents);
\item The \textit{duration of the counting period} for computing demand evolution (set to 60’);
\item The \textit{counting step} for computing demand evolution (set to 30’);
\item The \textit{maximum possible delay} (derived from the corresponding maximum delay imposed by the NM, as indicated in Table \ref{EvaluationCases});
\item The \textit{time horizon} \textbf{H} (here 24 hours);
\item The \textit{learning rate} $\alpha$ (set to 0.01 for all methods, as a “default” value although no such default exits);
\item The \textit{discount factor} $\gamma$ (set to 0.99 for all methods, as a “default” value although no such default exits);
\item The \textit{reward parameter} $\lambda$ (set to 20 for all methods, after experimentation made).
\end{itemize}

\section{Incorporating preferences on delays}

We have explored the capabilities of the MARL methods to solve DCB problems by incorporating strict conditions and preferences to the $MaxDelay$ for some of the flights. We denote the strict $MaxDelay$ imposed to a subset of flights with {\em Local MaxDelay}, while for the rest of the flights the $MaxDelay$ is as specified in Table \ref{EvaluationCases}. 

In doing so, we simulate situations where constraints of airlines to ground delays are specified and incorporated into the problem.  In these evaluation cases, the subset of flights is chosen according to the arrival airport: All flights arriving to one of the five biggest airports in Spain have a {\em Local MaxDelay} constraint, representing the need of less delay in airports that are hubs and with high traffic.
 
In these cases, roughly 30\% of the flights have a constraint on {\em Local MaxDelay},. These flights are also responsible to roughly the 30\% of the occurring hotspots. We considered sub-cases with {\em Local MaxDelay}, varying in {\em \{5, 10, 15, 25, 35, 45, 55\}}.

\begin{table*}[]
\caption{Average delay for the Jul2 case with various preferences on the local max delay.}
\begin{tabular}{|c|c|c|c|}
\hline
\textbf{Local MaxDelay} & \textbf{Average Delay(NM)} & \textbf{Average Delay(Ed-MARL)} & \textbf{Average Delay(IRL)} \\ \hline
\textbf{5}              & 1.66                       & 1.66                            & 1.76                        \\ \hline
\textbf{10}             & 1.66                       & 1.67                            & 1.77                        \\ \hline
\textbf{15}             & 1.66                       & 1.78                            & 1.81                        \\ \hline
\textbf{25}             & 1.66                       & 1.74                            & 1.78                        \\ \hline
\textbf{35}             & 1.66                       & 1.70                            & 1.80                        \\ \hline
\textbf{45}             & 1.66                       & 1.69                            & 1.79                        \\ \hline
\textbf{55}             & 1.66                       & 1.71                            & 1.81                        \\ \hline
\end{tabular}
\label{LocalMax_AvgDelay}
\end{table*}

\begin{table*}[]
\caption{Remaining hotspots and number of regulated flights for the Jul2 case with various preferences on the local max delay.}
\begin{tabular}{|c|c|c|c|c|}
\hline
\multirow{2}{*}{\textbf{Local MaxDelay}} & \multicolumn{2}{c|}{\textbf{Number of Resulting Hotspots}} & \multicolumn{2}{c|}{\textbf{Number of Regulated Flights}} \\ \cline{2-5} 
                                         & \textbf{Ed-MARL}               & \textbf{IRL}              & \textbf{Ed-MARL}              & \textbf{IRL}              \\ \hline
\textbf{5}                               & 3                              & 2                         & 402.2                         & 430.4                     \\ \hline
\textbf{10}                              & 1                              & 0                         & 408.7                         & 461.7                     \\ \hline
\textbf{15}                              & 0                              & 0                         & 432.4                         & 461.7                     \\ \hline
\textbf{25}                              & 0                              & 0                         & 448.2                         & 460.2                     \\ \hline
\textbf{35}                              & 0                              & 0                         & 442.6                         & 462.2                     \\ \hline
\textbf{45}                              & 0                              & 0                         & 438.2                         & 465                       \\ \hline
\textbf{55}                              & 0                              & 0                         & 439.1                         & 466.5                     \\ \hline
\end{tabular}
\label{LocalMax_Flights}
\end{table*}

We present results for the most difficult case, among the cases considered: Jul2. Tables \ref{LocalMax_AvgDelay} and \ref{LocalMax_Flights} show that MARL methods manage to solve DCB problems using the individual delay reward, even if very strict constraints on {\em Local MaxDelay} are set to a subset of the flights. However, for {\em Local MaxDelay = 5’}, both methods could not provide a solution and for {\em Local MaxDelay = 10’} only IRL could resolve all hotspots. 

In the cases were both methods succeed to provide solutions, the Ed-MARL approach provides the lower average delay per flight, and the lower number of regulated flights, consistently with the results already provided in previous sections. It must be noticed, that the results of all MARL methods are close to the results obtained when all flights adhere to the same $MaxDelay$.

\end{document}